\documentclass[12pt]{article} %***
\usepackage[sectionbib]{natbib}
\usepackage{array,epsfig,fancyheadings,rotating}
\usepackage[]{hyperref}  %<----modified by Ivan
\usepackage{sectsty, secdot}
\sectionfont{\fontsize{12}{14pt plus.8pt minus .6pt}\selectfont}
\renewcommand{\theequation}{\thesection\arabic{equation}}
\subsectionfont{\fontsize{12}{14pt plus.8pt minus .6pt}\selectfont}
\usepackage{amsmath}
\usepackage{amssymb}
\usepackage{amsfonts}
\usepackage{multirow}
\usepackage{amsthm}
\newtheorem{corollary}{Corollary}

\theoremstyle{definition}

\usepackage{setspace}
\usepackage{color}
\usepackage{multirow}
\usepackage{xcolor}
\usepackage{bm} 
\usepackage{enumitem}
\newtheorem{thm}{Theorem}
\newtheorem{prop}{Proposition}
\numberwithin{equation}{section}

\def \cY{\bm{Y}}
\def \cW{\bm{W}}

\newcommand{\beq}[1]{\begin{equation}\label{#1}}
\newcommand{\eeq}{\end{equation}}

\newcommand{\paren}[1]{\left(#1\right)}

\newcommand{\mb}[1]{\boldsymbol{#1}}

\newcommand{\E}[1]{\mathbb{E} \paren{#1} }
\newcommand{\V}[1]{\mathbb{V}\paren{#1}}

\newcommand{\Prob}[1]{\mathbb{P}\paren{#1}}

\renewcommand{\arraystretch}{1.5}

\newcommand\independent{\protect\mathpalette{\protect\independenT}{\perp}}
\def\independenT#1#2{\mathrel{\rlap{$#1#2$}\mkern2mu{#1#2}}}

\begin{document}

%%%%%%%%%%%%%%%%%%%%%%%%%%%%%%%%%%%%%%%%%%%%%%%%%%%%%%%%%%%%%%%%%%%%%%%%%%%%%%%%%%%%%%%%%%%%%%%%%%%%%%%%%%%%%%%%%%%%%%%%%%%%
%%%%%%%%%%%%%%%%%%%%%%%%%%%%%%%%%%%%%%%%%%%%%%%%%%%%%%%%%%%%%%%%%%%%%%%%%%%%%%%%%%%%%%%%%%%%%%%%%%%%%%%%%%%%%%%%%%%%%%%%%%%%

\renewcommand{\baselinestretch}{2}

\markright{ \hbox{\footnotesize\rm Statistica Sinica
%{\footnotesize\bf 24} (201?), 000-000
}\hfill\\[-13pt]
\hbox{\footnotesize\rm
%\href{http://dx.doi.org/10.5705/ss.20??.???}{doi:http://dx.doi.org/10.5705/ss.20??.???}
}\hfill }

\markboth{\hfill{\footnotesize\rm FIRSTNAME1 LASTNAME1 AND FIRSTNAME2 LASTNAME2} \hfill}
{\hfill {\footnotesize\rm FILL IN A SHORT RUNNING TITLE} \hfill}

\renewcommand{\thefootnote}{}
$\ $\par

%%%%%%%%%%%%%%%%%%%%%%%%%%%%%%%%%%%%%%%%%%%%%%%%%%%%%%%%%%%%%%%%%%%%%%%%%%%%%%%%%%%%%%%%%%%%%%%%%%%%%%%%%%%%%%%%%%%%%%%%%%%%

\fontsize{12}{14pt plus.8pt minus .6pt}\selectfont \vspace{0.8pc}
\centerline{\large\bf Propensity score regression for causal inference}
\vspace{2pt} 
\centerline{\large\bf  with treatment heterogeneity}
\vspace{.4cm} 
\centerline{Peng Wu$^{a,b}\dag$, Shasha Han$^{c,d}\dag\footnote{\dag contributed equally.}$, Xingwei Tong$^b$ and Runze Li$^{e}\ast$ \footnote{$\ast$correspond to: rzli@psu.edu}} 
\vspace{.4cm} 
\centerline{\it $^a$Beijing Technology and Business University, 
$^b$Beijing Normal University,}  
\centerline{\it  
$^c$Chinese Academy of Medical Sciences and Peking Union Medical College,}
\centerline{\it 
$^d$Peking University, and 
$^e$The Pennsylvania State University}
 \vspace{.55cm} \fontsize{9}{11.5pt plus.8pt minus.6pt}\selectfont

%%%%%%%%%%%%%%%%%%%%%%%%%%%%%%%%%%%%%%%%%%%%%%%%%%%%%%%%%%%%%%%%%%%%%%%%%%%%%%%%%%%%%%%%%%%%%%%%%%%%%%%%%%%%%%%%%%%%%%%%%%%%

\begin{quotation}
\noindent {\it Abstract: Understanding how treatment effects vary on several key characteristics is critical in the practice of personalized medicine. In such cases, nonparametric estimation of these conditional average treatment effects is often desirable. However, few 
 methods are available owing to the computational difficulty of such estimations. Furthermore, existing nonparametric methods, such as the inverse probability weighting methods, have limitations that hinder their use when the values of propensity scores are close to zero or one. We propose a propensity score regression (PSR) method that allows nonparametric estimation of such conditional average treatment effects in a wide context. The PSR comprises two nonparametric regressions. First, it regresses on the propensity scores together with the characteristics of interest, to obtain an intermediate estimate. Then, it regresses the intermediate estimate on the characteristics of interest only. By including propensity scores as regressors in a nonparametric manner, the PSR eases the computational difficulty substantially while remaining less sensitive to the values of propensity scores. We present its several appealing properties, including consistency and asymptotical normality. In particular, we show the existence of an explicit variance estimator, which we use to assess the analytical behavior of the PSR and its precision.  
  The results of our simulation studies indicate that the PSR outperforms existing methods in various settings with extreme values of propensity scores. We apply our method to the national 2009 flu survey (NHFS) data to investigate the effects of seasonal influenza vaccinations and having paid sick leave across different age groups.}

\vspace{9pt}
\par 
\end{quotation}\par

\def\thefigure{\arabic{figure}}
\def\thetable{\arabic{table}}

\renewcommand{\theequation}{\thesection.\arabic{equation}}

\fontsize{12}{14pt plus.8pt minus .6pt}\selectfont

% =========================================================================
% Section 1
\section{Introduction}
\label{sec:intro}

The heterogeneous treatment effect describes the effect variability due to varying characteristics and is widely used in contexts such as personalized medicine, policy design, and customized marketing \citep{Kent-etal2018, Yin2018, Imai-Strauss2011, Sato-etal2019}. In many settings, the characteristics of treatment relevance are only a subset of the baseline covariates ($X = (X^l, X^{-l})$). Understanding how a treatment works for individuals that differ on these few core characteristics ($X^l$) is particularly critical for developing tailored treatment decisions. For example, in clinical practice, older patients tend to suffer more from side effects or drug-drug interactions. Thus, the age-dependent drug efficacy of a treatment is used to balance its benefits and risks \citep{Velentgas-etal2013}. In health policy, age-dependent vaccine effectiveness is used to guide targeted vaccination programs \citep{Soiza-etal2021}. Compared with the conditional treatment effects given the full covariates \citep{Nie-Wager-2018, Wager-Athey-2018}, these conditional treatment effects given the key characteristics are more easily interpretable and are widely used in clinical settings.

However, estimating such conditional treatment effects is challenging, because such methods should be able to flexibly distinguish the heterogeneous effect (which is conditional on $X^l$) from the effect due to the remaining confounding covariates ($X^{-l}$). Because $X^{-l}$, which can be high-dimensional, may still confound the effects of treatments on outcomes, conditioning on $X^l$ is not sufficient. Moreover, the degree of confounding may vary with $X^l$, making modeling the conditional outcome particularly challenging. A nonparametric estimation method allows a fully flexible model and is therefore desirable. However, few nonparametric methods for estimating these heterogeneous treatment effects are available, and the weighting-based methods have limitations that hinder their use in a wide context. For example, the inverse probability weighting proposed by \cite{Abrevaya-Hsu-Lieli-2015} uses the inverse of propensity scores as weights to adjust outcomes. However, this method can result in unstable estimates when the values of propensity scores are close to zero or one--- that is, the weights are very large, as typically observed in weighting methods to population average treatment effects \citep[see e.g.,][]{Hahn-1998, Rubin-2001, Kang-Schafer-2007}.  The augmented inverse probability weighting (AIPW) methods \citep{Lee-Okui-Whang-2017} also use the inverse of propensity scores as weights, and they require correctly parametric modeling outcomes to achieve efficiency \citep{Seaman-Vansteelandt2018}. Although the requirement can be relaxed by leveraging machine learning methods \citep{Fan-Hsu-Lieli-Zhang-2019, Zimmert-Lechner-2019, Semenova-Chernozhukov}, these methods can rely heavily on extrapolation, which is a critical concern in the context with extreme propensity score values  \citep[]{Kang-Schafer-2007, Tan2007, Wu-etal-Bio2022}.

In general, alternative methods rely on a two-step estimation: first, they estimate the conditional treatment effects defined on the full covariates, and then they integrate out the obtained estimates to the desired level of granularity.  However, it is often difficult to estimate the conditional treatment effect nonparametrically for a high-dimensional covariate \citep{Abrevaya-Hsu-Lieli-2015, Lechner-2019, Zimmert-Lechner-2019, Wu-Tan2021}. For example, in our example based on data from a national 2009 flu survey (NHFS)  (see \S \ref{real example}), the dimension of the full covariates is as high as 65 (see Supplementary Material). With such a high dimension, typical nonparametric estimation methods, for example, the local linear regression, would suffer from the so-called curse of dimensionality \citep{Fan-Gijbels1996}.

Following a two-step estimation, we propose a nonparametric propensity score regression (PSR) method, consisting of two nonparametric regressions. First, it regresses the propensity scores together with the covariates of interest. Then, it integrates out the scores by regressing the estimates from the first regression on the covariates of interest only. Propensity scores exhibit a crucial balancing property, namely, the distributions of full covariates between the treatment groups are identical at each level of the propensity scores (including the one-to-one functions of propensity scores). In our context, the balancing property is useful for controlling the confounding due to the remaining covariates $X^{-l}$, and for easing the computational difficulty. The PSR uses a continuous and bounded function of the score, and therefore is less sensitive to extreme propensity scores.  
 Furthermore, by using the propensity scores in a nonparametric manner in the first step, the PSR reduces  
% \bcol{smoothes}  
  the influences of errors in the propensity scores on the estimates of the second step \citep{Mammen-etal2012}, and thus enjoys increased robustness to such errors. On the other hand, weighting-based  methods achieve a covariate balance for a hypothetical super-population constructed by reweighing units in the study population, where small changes in propensity scores could lead to large discrepancies in weights, and even nonparametric estimation can result in highly unstable estimates.

%  Hahn-Ridder2013,  
The idea of including propensity scores in regression is not new in the context of parametric estimators of average treatment effects \cite[see e.g.,][]{Little-An-2004, Zhang-Little-2009, Zhou-Elliott-Little-2019, Wu-etal-2021}. However, these approaches still rely on technical modeling assumptions for the outcome. When using propensity scores in a parametric regression, the key is to correctly specify the elusive relationship between the propensity score and the outcome, which is intrinsically connected to difficulties in specifying outcome models.  Unlike these methods, we propose using propensity scores as regressors in a nonparametric manner in order to estimate heterogeneous treatment effects. 
% \cite{Zhang-Little-2009} proposed to use a spline of the logit of the propensity in regression for flexible functional forms between the outcome and the propensity score. 
 % \pw{TODO: some references of Hahn..}
 
 We validate the approach theoretically and show its appealing advantages. To obtain the theoretical results, we assume a parametric estimation of propensity scores, but do allow a nonparametric estimation of propensity scores. Note that, even under the parametric assumption, unlike the weighting-based methods, the PSR allows a one-to-one transformation of propensity scores, and the functional form of the propensity score is less important. We present the theoretical properties of the proposed method, including the consistency and asymptotical normality, and an explicit variance estimator, which we use to assess the 
  analytical behavior of the PSR and its precision.

The PSR is not only valuable for exploring treatment heterogeneity for practical guidance, but also useful in understanding treatment heterogeneity with high-dimensional full covariates. For example, we can decompose the full covariates into many subsets, each with only a few covariates, and estimate the heterogeneous treatment effects on these subsets. With this ensemble of such heterogeneous treatment effects, we may approach the full picture of the treatment heterogeneity with high-dimensional full covariates, which is computationally difficult to estimate directly \citep{Abrevaya-Hsu-Lieli-2015, Lechner-2019, Zimmert-Lechner-2019, Wu-Tan2021, Semenova-Chernozhukov}. 

 The remainder of the paper proceeds as follows. In \S \ref{motivations and literatures}, we introduce the basic framework and the motivation of the analysis. In \S \ref{PSR}, we present the PSR method. Here, \S \ref{PSR: description} outlines the method and provides the theoretical validation, and \S \ref{PSR: estimator} provides the nonparametric estimator. In \S \ref{PSR: properties}, we show the theoretical properties. We conduct several simulation studies in \S \ref{sec: sim}. In \S \ref{real example}, we apply our method to the national 2009 flu survey (NHFS) data to investigate the effects of seasonal influenza vaccination and having paid sick leave across different age groups. We conclude with a discussion in \S \ref{sec: discussion}. 
% All technical proofs are given in the Supplementary Material. finite-sample 

% ===========================================================================
\section{Motivations}\label{motivations and literatures}
\subsection{Notation and assumptions}

We adopt the framework of the Rubin Causal Model \citep{Rubin1974}, also called the potential outcome approach to causal inference \citep{Imbens-Rubin-2015}. Consider a study with $N$ units. Each unit $i = 1, \ldots, N$ is associated with a vector-valued covariate $X_i = (X_{i1},  \ldots,X_{il},\ldots, X_{ip}) \in \mathcal{X} \subseteq \Re^p$, measured before being exposed to  treatment $D_i$. The low-dimensional covariates of interest is denoted by $X_i^l = (X_{i1}, \ldots,X_{il}) \in \mathcal{X} ^l \subset \Re^p$. We write $X_i = (X_i^l, X_i^{-l})$. The outcome variable $Y$ is measured on each unit after its treatment exposure. Associated with treatment $d$, $d = 0, 1$, is the potential outcome $Y_i(d)$, the value of $Y$ when unit $i$ is exposed to treatment $t$, which implicitly assumes the stable unit treatment value assumption \citep[SUTVA, ][]{Rubin-1980}.  When referring to a generic unit, we drop the subscript and write
$X,D,Y (0),Y (1),Y, X^l, X^{-l}$, and so on.

As in the literature \citep[e.g.,][]{Abrevaya-Hsu-Lieli-2015}, we assume the unconfounded assumption, namely, 
% the treatment assignment is unconfounded given the covariates $X$, 
%\begin{equation} \label{UncoAssump}
$ D \independent (Y(1), Y(0)) ~| X.$ 
% \end{equation} 
We denote the propensity score $e(X) : = \Prob{D = 1 | X}$, assuming that $0< e(x) < 1$, for any $x \in \mathcal{X}$.  The heterogeneous treatment effect of interest, $\tau(x^l)$, is defined on the subspace of the covariates, $\mathcal{X}^l$, as
\begin{equation} \label{eq:HTE}  
         \tau(x^l)  := \E{ Y(1) - Y(0) | X^l = x^l },
\end{equation}
where $x^l \in \mathcal{X}^l$. Often, the dimension of $\mathcal{X}^l$ is much smaller than that of the full covariates space $\mathcal{X}$. 
Therefore, $\tau(x^l)$ is at a higher level of granularity than the treatment effects conditional on the full covariates.

\subsection{Two-step estimation}

To estimate an estimand at a higher level of granularity, an intuitive way is to estimate the treatment effects at a lower level of granularity first, and then integrate out the obtained estimates into the subspace of interest. Our idea is broadly a two-step estimation. The key intuition is to explore an estimand at a lower level of granularity that can be estimated unbiasedly and nonparametrically. Below, we describe this idea further.

% To \pw{start}, 
We write $\tau(x^l)$ using the tower property of conditional expectation as
$$
\begin{aligned}
 \tau(x^l)  
 &= \mathbb{E} \left [ \E{ Y(1) - Y(0) | X^l = x^l, X^{-l} } | X^l = x^l \right ].\\
%&= \E{ \eta(x^l,X^{-l} )}
\end{aligned}  
$$   
Let  $\eta(x)  := \E{ Y(1) - Y(0) | X^l = x^l, X^{-l} = x^{-l}}$, where $x= (x^l, x^{-l})$. In principle, we can estimate the insider expectation $\eta(X)$ first, and then integrate $X^{-l}$ out with respect to the conditional distribution of $ X^{-l}$ given $X^l = x^l$. In this case, our task is to estimate the finest estimand $ \eta(x)$ for each $x \in \mathcal{X}$, which can be identified as
%   Under unconfoundedness assumption, % \eqref{UncoAssump}, 
 % $\eta(x)$ can be identified as  
{\small \begin{equation}\label{eq: identification hte full}
\begin{aligned}
\eta(x) 
 &= \E{Y(1) | D = 1, X^l = x^l, X^{-l} = x^{-l} } - \E{Y(0) | D = 0, X^l = x^l, X^{-l} = x^{-l}} \\
&= \E{ Y| D = 1, X^l = x^l, X^{-l} = x^{-l} }- \E{ Y | D = 0, X^l = x^l, X^{-l} = x^{-l} }.
\end{aligned}  
\end{equation}}
However, when the dimension $p$ of the full covariates space is large, nonparametric estimation of $\eta(X)$ can be difficult \citep{Abrevaya-Hsu-Lieli-2015, Lechner-2019, Zimmert-Lechner-2019, Wu-Tan2021}.

This motivates us to explore an alternative estimand for the first-step estimation.  Specifically, we aim to find an estimand that lies in a much higher level of granularity than $ \eta(X)$ while still lying in a lower level than $\tau(X^l)$, so that in practice it is possible to estimate the new estimand nonparametrically. Notably, the lower auxiliary variable needs to replace the important role that $X^{-l}$ plays in the identification. As illustrated in Equation \eqref{eq: identification hte full}, conditioning on $X^{-l}$ and $X^l$, i.e., the full covariates $X$, facilitates identifying $\eta(X)$ using the observed data, owing to the natural balancing property of the full covariates $X$. 
 
 The new auxiliary variable needs to rest on a subspace with a dimension much smaller than $p$, and should exhibit the aforementioned balancing property.   
The propensity score, defined as the probability of assignment to the treatment given the full covariates \citep{Rosenbaum-Rubin-1983}, is one candidate.  As a summary of covariates, the propensity score reduces the $p$-dimensional covariates into a scaler while exhibiting the desired balancing property. Clearly, any one-to-one functions of the propensity score are candidates as well.

\subsection{Comparison with existing methods using propensity scores}
% Existing methods have leveraged propensity scores to achieve covariate balance, but for a hypothetical super-population which is constructed from reweighing the units of the studying population.
Existing methods use propensity scores to achieve the covariate balance by reweighing the units.  
 For example, \citet{Abrevaya-Hsu-Lieli-2015} proposed the inverse probability weighting (IPW) estimator % nonparametric 
 \begin{equation} \label{eq:IPW}
 \tau^{IPW} (x^l) =  \mathbb{E} \Big (  DY/e(X) - (1-D)Y/(1-e(X))   \Big | X^l = x^l   \Big ).   
 \end{equation} 
% parametric nonparametric 
 As in the weighting methods \citep[see e.g.,][]{Hahn-1998, Rubin-2001, Kang-Schafer-2007}, the IPW estimator in (\ref{eq:IPW}) is sensitive to the estimated propensity score values and their estimates are highly unstable if the propensity score values are close to zero or one.  
When the parametric function of the outcome model is knowable, \citet{Lee-Okui-Whang-2017} propose the AIPW estimator 
{\footnotesize \begin{equation} \label{eq:AIPW}
\tau^{AIPW} (x^l)  =  \E{  \frac{D\paren{Y - \mu_1(X)}}{e(X)} - \frac{(1-D)\paren{Y - \mu_0(X)}}{1-e(X)} +  \paren{ \mu_1(X) - \mu_0(X)} \bigg | X^l = x^l    },    
 \end{equation}}
 where $\mu_d(\cdot)$ for $d =0,1$ are specified outcome functions.
 The AIPW allows for misspecification of the propensity score model if the parametric outcome functions $\mu_d(\cdot)$ are specified correctly. However, with high-dimensional covariates, correct specifications of outcome functions $\mu_d(\cdot)$ are not easy. Several prior works have tried estimating  $\mu_d(\cdot)$ using machine learning methods \citep[e.g.,][]{Fan-Hsu-Lieli-Zhang-2019, Semenova-Chernozhukov}, but these methods rely heavily on  extrapolation \citep[e.g.,][]{Kang-Schafer-2007, Tan2007}.   % Literatures have instead explored techniques to estimate

Like the IPW,  the AIPW is also sensitive to the extreme propensity score values, even when the propensity score models are specified correctly \citep{Rotnitzky-Vansteelandt2014}. 
Instead of using propensity scores as weights, we include them as one regressor in nonparametric regression, yielding the PSR method. 
% We call the generic method propensity score regression (PSR). 

% It is important to note that 
The PSR is conceptually different from the propensity score weighting to achieve the balance. The weighting methods use the inverse of propensity scores as weights to construct a hypothetical super-population, in which the distributions of the covariates between the treated units and the control units can be balanced. Clearly, these weights are unbounded around zero or one, with small changes in propensity scores leading to potentially large discrepancies in weights, particularly when some  
 propensity score values are extreme. The current setting is different. Here, $\beta(X^l,e)$ defined in equation (\ref{beta1}) is a bounded function of the propensity score, and is therefore less sensitive to extreme values of the propensity score.  

As such, the PSR is analogous to propensity score matching and subclassification, all of which are based on the balancing property of propensity scores in the study population. But unlike matching on propensity scores, which implicitly involves model specifications (e.g., we need to specify the matching criteria and, in general, different criteria lead to different matched sets), the PSR uses propensity scores in a nonparametric manner, that is, using existing nonparametric methodologies. With parametric modeling on propensity scores, the difficulty of matching extreme propensity score values is likely to lead to substantial bias. However, here the estimation is nonparametric and  $ \beta(x^{l}, e)$ is estimated smoothly. Thus, the results should be less sensitive to minor differences between inexact matches. We show this using a variant of the PSR in which the regression procedure is replaced by matching on propensity scores (\S \ref{sec: sim}). 

Note that with large differences between the propensity scores of the treated units and the control units --- for treated units near a propensity of 1.0 \emph{only}, and for control units near a propensity of 0.0 \emph{only} --- the PSR may not be suitable either. 
 We note that in a mild situation, where both some treated units and control units near a propensity of 1.0 and 0.0, IPW and AIPW estimators could generate highly sensitive estimates. % \pw{Should we need this paragraph?}  

  \section{PSR}\label{PSR}  
The PSR includes two nonparametric regressions as follows. 
% We detail them below in turn.   
For notational simplicity, we refer to ``$e$'' as the value of the propensity score $e(X)$ or any one-to-one function of $e(X)$.  As discussed in  \S \ref{motivations and literatures}, the key idea is to explore an intermediate estimand at a lower level of granularity that can be estimated unbiasedly and nonparametrically. We denote $\beta(X^{l}, e)$ as the intermediate estimand, which is conditional on the propensity score and the covariates of interest, as  
\begin{equation} \label{beta1} 
\beta(x^{l}, e) = \E{ Y(1) - Y(0)| X^{l} = x^l, e(X) = e }.\\
\end{equation}

%We would like to note 
Note that the estimand $\beta(x^{l}, e )$ is conditional on the $(l+1)$-dimensional variables, where $l+1$ is substantially smaller than the dimension of the full covariates $p$, 
% and, unlike $\eta(x)$, it has only one extra variable, the propensity scores,  substantially
 and therefore can mitigate the problem of high-dimensional covariates.  
Estimating $\beta(x^{l}, e )$ is our central task. Note that the definition of $\beta(x^{l}, e )$ includes both potential outcomes $Y(1)$ and $Y(0)$, but only one of them is observed in the real world. By conditioning on the propensity score, we can replace the potential outcomes with the observed outcome $Y$.  Below, we prove this nonparametrically.

\begin{prop} \label{prop: local reg} 
Suppose that $\E{Y | D,  X^{l}, e }$ is a nonparametric regression function. Then, $\beta(X^{l}, e )$ and $ \E{ Y(0) | X^{l}, e}$ are the functional coefficients corresponding to the treatment indicator $D$ and the intercept, respectively:  
\begin{equation}  \label{eq:HTE-model}
 \E{ Y | D,  X^{l}, e } =  \beta(X^{l}, e )\cdot D + \E{ Y(0) | X^{l}, e },
  \end{equation}
  where $\beta(x^{l}, e) = \E{ Y \mid D=1, X^{l} = x^{l}, e(X) = e }-  \E{  Y \mid D=0, X^{l} = x^{l}, e(X) = e }$.  
\end{prop} 

% Fan-Zhang-1999, Fan-Zhang-2008
Model (\ref{eq:HTE-model}) is a varying coefficient model \citep{Hastie-Tibshirani-1993} and can be estimated using standard local linear regression techniques \citep{Fan-Zhang-1999}. Note that we do not make any parametric assumptions. The model is general enough to capture any model specification. The counterpart working model is given by
\begin{equation}   \label{beta working} 
Y  =   \beta(X^{l}, e)\cdot D  +   \E{ Y(0) | X^{l}, e }  + \xi,
\end{equation} 
where $\E{\xi | D, X^{l},  e} = 0$. 

Remarkably,  here $\beta(x^{l}, e )$ is estimated using the full sample of data under both two treatment conditions, $d= 0, 1$. Unlike existing methods such as the AIPW, our approach does not extrapolate the unobserved outcomes using predicted values from the other treatment group. Consequently, it is more likely to exhibit good finite-sample performance.  
 %  it  has no problem of making implicit extrapolation, and may have better finite sample performance than the method that splitting the sample into treated group and control group.   
 Once we have estimates of $\beta(x^{l}, e)$, we simply integrate out the propensity scores $e$ to obtain the estimates for $\tau(x^l)$. To do so, we conduct a second nonparametric estimation based on the projection relationship of $\tau(x^l)$ and $\beta(x^{l}, e)$. 
\begin{prop} \label{prop: local prob} 
$ \tau(x^l)$ is, geometrically, the projection of $\beta(X^{l}, e)$ into the subspace spanned by $X^l$, specifically, $\tau^{PSR}(x^l)  = \E{ \beta(X^{l}, e ) | X^{l} = x^l }.$
 % \sum_{\beta} \beta \cdot \frac{\Prob{\beta(X^{l}, e ) = \beta , X^{l} = x^l}}{\Prob{X^l = x^l}}.
 % \]
\end{prop}

Proposition \ref{prop: local prob} suggests that $ \tau(x^l)$ can be estimated nonparametrically, for example, using a local linear regression of $\beta(X^{l}, e)$ on $X^l$.

 \subsection{Description of the approach}  \label{PSR: description}
 When $e(X)$ is known, the PSR is implemented using two nonparametric regressions, built on Propositions \ref{prop: local reg} and \ref{prop: local prob}, respectively.   
When $e(X)$ is also estimated from the data, the PSR is implemented in a total of three steps:   
\begin{description} % the values of 
\item[Step 0:] Estimate $e(X)$ in either a parametric or a nonparametric manner.
\item[Step 1:] Estimate  $\beta(X^{l}, e)$ by nonparametrically regressing the outcome $Y$ on the covariate $X^l$ and the propensity scores $e(X)$.
\item[Step 2:] Estimate $\tau(X^l)$ by nonparametrically regressing the estimated values of $\beta(X^{l}, e)$ on the covariate $X^l$ only.
\end{description}

% We propose this three-step PSR approach with propensity scores estimated either parametrically or nonparametrically. Acknowledgedly, the nonparametric estimation of propensity scores $e(X)$ may have the same computational problems as that in the nonparametric estimation of the treatment effects conditional on the full covariate, $\eta(X)$. We illustrate the use of this three-step PSR approach in several simulation studies.

For practical use, we are interested in the large-sample properties of $\hat \tau(x^{l})$. Here, we need to determine how 
  the errors of the estimated propensity scores in Step 0 affect the estimates for $\tau(x^l)$. Briefly, the PSR is robust to estimation errors of the propensity scores, provided that the influence of these errors has a negligible effect on the second step estimation \citep{Mammen-etal2012}. In our context,   this is because the propensity scores are used in a nonparametric manner (Step 1), where the true propensity scores and their estimated counterparts are asymptotically indistinguishable.

Below, we focus on continuous $X^{l}$, but our results hold in the general setting including discrete $X^{l}$. 
 For additional details on the discrete $X^{l}$, please refer to \S 7 of the Supplementary Material. % (\bcol{Note I move the comment here.})
% Readers who are interested in the details for 

\subsection{Nonparametric estimator}\label{PSR: estimator}
% In the section, 
%We present the nonparametric estimators of $\beta(x^{l}, e)$ and $\tau(x^l)$ in Steps 1 and 2 respectively. %In \pw{Steps} 1 and 2, 
We use the standard local linear regression as the nonparametric method to estimate $\beta(x^{l}, e)$ and $\tau(x^l)$ in Steps 1 and 2, respectively.  For notational simplicity, we focus on the case of $l = 1$ and consider a more general case in the Supplementary Material.

We consider the case in which the propensity scores are known, and denote the response vector  $\bm{Y} = (Y_{1}, \cdots, Y_{N})^{\intercal}$,  the regressor vector $\bm{\Gamma} = (\Gamma_{1}, \cdots, \Gamma_{N})^{\intercal}$, with the regressors  $\Gamma_{i} =  (D_{i},
1, D_{i}(X^l_{i} - x^l)/h_{1},(X^l_{i} - x^l)/h_{1}, D_{i} ( e_{i} - e)/h_{2}, ( e_{i} - e)/h_{2})^{\intercal}$ for $i = 1, \ldots, N$, and the kernel vector $\bm{W} = \text{diag} \{ K_{h_{1}}(X^l_{1}-x^l)K_{h_{2}}( e_1- e), \ldots, K_{ h_{1} }(X^l_{N}-x^l) K_{h_{2}}( e_N-e) \}$, with % the kernel functions  
$K_{h_{j}}(u) = K(u/h_{j}) /h_{j}$ for $j=1, 2$. % \pw{$h1$ may be a vector..} 
The standard local linear estimator $\tilde{\beta}(x^l, e)$ of $\beta(x^l, e)$ is
%  \pw{the first component of}  
 \begin{equation} \label{tilde-beta}  
 \tilde{\beta}(x^l, e) =     (1, 0, 0, 0, 0, 0)
  (\bm{ \Gamma}^{\intercal} \bm{ \cW}\bm{\Gamma})^{-1} \bm{\Gamma}^{\intercal} \bm{\cW \cY},    
 \end{equation}
 % that is, $\tilde{\beta}(x^l, e)$ corresponds to
which is  the first component of $ (\bm{ \Gamma}^{\intercal} \bm{ \cW}\bm{\Gamma})^{-1} \bm{\Gamma}^{\intercal} \bm{\cW \cY}$. 
% where  \rcol{$I_{2}$ is an identity matrix with size $2$, $0_{2\times 4}$ is a size $2 \times 4$ matrix with each elements being 0. } 
Furthermore, we denote that the response vector $\bm{\tilde{\beta}} =  ( \tilde{\beta}(X^l_1, e_1), \ldots, \tilde{\beta}(X^l_N, e_N) )^{ \intercal }$, the regressor vector $\bm{G} = (G_{1}, \ldots, G_{n})^{\intercal}$ with $G_{i} = (1, (X^l_{i}- x^l)/h_{3})^{\intercal}$, and $\bm{\Lambda} 
=\text{diag}\{ K_{h_{3}}(X^l_{1}- x^l), \ldots,  K_{h_{3}}(X^l_{N}- x^l)\}$.
The local linear estimator of $\tau(x^l)$ is  
 \begin{equation}\label{tilde-tau}
 \tilde{\tau}(x^l) = (1, 0) (\bm{G}^{\intercal} \bm{\Lambda} \bm{G} )^{-1} \bm{G}^{\intercal} \bm{\Lambda}  \tilde{\mb{\beta}}.
\end{equation}  

Next, we consider the case where the propensity scores are estimated from the data. We replace $\bm{ \Gamma}$,$\bm{\cW}$, and $ \tilde{\bm{\beta}}$ in \eqref{tilde-beta} and \eqref{tilde-tau} with $\hat{\bm{\cW}}$, $\hat{\bm{ \Gamma}}$, and $ \hat{\bm{\beta}}$, respectively. Note that we replace the true propensity scores $e$ by the estimated $\hat{e}$. Using the estimated propensity scores, the local linear estimator of $\beta(X^{l}, e)$ and $\tau(x^l)$ is then given by
\begin{equation} \label{hat-beta} 
\hat{\beta}(x^l, e) =  (1, 0, 0, 0, 0, 0)  (\hat{\bm{ \Gamma}}^{\intercal}\hat{ \bm{ \cW}} \hat{\bm{\Gamma}})^{-1} \hat{\bm{\Gamma}}^{\intercal} \bm{\hat{\cW} \cY},    
 \end{equation} 
 \vskip -1cm
 \begin{equation}\label{hat-tau}
 \hat{\tau}(x^l) = (1, 0) (\bm{G}^{\intercal} \bm{\Lambda} \bm{G} )^{-1} \bm{G}^{\intercal} \bm{\Lambda  \hat{\beta}}.
\end{equation}

% ======================================================================
% ======================================================================
\subsection{Theoretical properties}\label{PSR: properties} 
%To answer the question proposed in \S \ref{PSR: description}, 
We first present the following key assumption. 

\noindent 
\emph{Assumption 1.} The propensity score model can be written as  $e(X) = g(X^\intercal \alpha)$, where $\alpha$ is the true unknown parameter, and $g(\cdot)$ is a known function (e.g., generalized linear model). 
	\begin{itemize}
		\item[(i)] The estimates of $\alpha$, denoted by $\hat \alpha$, satisfies $\hat \alpha - \alpha  = O_{p}(N^{-1/2})$;  
	      \item[(ii)]  The second-order derivative of $g$ is uniformly bounded, that is, $\sup_{t} |g''(t)| $ is bounded.
	\end{itemize}	

Assumption 1(i) is critical to our main results because it simplifies the asymptomatic analysis on $\hat{\tau}(X^l)$. Note that the parametric modeling assumption on propensity scores can be relaxed, see the Discussion for more details.
  
\begin{thm} \label{theorem: beta}  Under Assumption 1 and regularity Assumption 2 in the
  Supplementary Material, $\hat \beta(x^{l}, e)$ in \eqref{hat-beta}  and  $\tilde \beta(x^{l}, e)$ in \eqref{tilde-beta} are asymptotically indistinguishable, that is, 
$
           \sup_{ \paren{x^l, e} \in \mathcal{X}^l \times(0,1) } \big | \hat \beta (x^l, e) - \tilde \beta (x^l, e) \big |  =  O_{p}(N^{-1/2}).     
$
\end{thm} 

Theorem \ref{theorem: beta} demonstrates that the effect of the estimation error of the propensity score on the estimator of $\beta(x^{l}, e)$ is negligible. Intuitively, this is because the propensity score is used in a nonparametric manner in Step 1, where we need only locally indistinguishable estimates of the propensity scores from the true values. This finding is consistent with the results of \citep{Mammen-etal2012} in a different setting on studying the coverage rates of the final estimates. The latter study shows that the effect of the first-step estimation error on the second-step estimation is restricted in a smoothed way through the estimation bias in the first step.  From Theorem \ref{theorem: beta}, the effect on the estimator of $\tau (x^l)$ is small \cite[e.g.,][]{Gu-Yang-2015}. We next establish the asymptotical normality of $\hat \tau(x^{l})$. Let $\bar{K}(x) = \int K(t)K(x + h_{1}t/h_{3} )dt$, $\nu= \int K^{2}(t)dt$, where $f(x^l)$ is the density function for $X^l$.
\begin{thm} \label{th2} 
Under Assumption 1 and  regularity Assumptions 2--3 in  the Supplementary Material,  $\hat \tau (x^l)$ in \eqref{hat-tau} is a consistent estimator, and  %  of $ \tau (x^l)$ 
% in \eqref{eq:HTE} and  
$$ 
\V{x^l}^{-1/2}  \paren{  \hat \tau (x^l) - \tau(x^l)  }     \xrightarrow{d}  N(0, 1), 
$$
where $\V{x^l}$ represents the asymptotic variance and
{\small 
$$
  \V{x^l}  = \frac{1}{Nh_3 f(x^l) } \paren{  \nu \cdot \V{ \beta(X^{l}, e) | X^l = x^l}  +  \int \bar{K}^{2}(t)dt \cdot   \E{  \frac{ (D - e)^{2}}{e^{2} (1 - e)^{2}}  \xi^{2} | X^l = x^l }  }.
$$ }
  \end{thm}

Note that Theorem 2 relies on Assumption 3(iv) $\sqrt{N h_3} ( h_{1}^{2} + h_{2}^{2} + h_{3}^{2} ) \to 0$ as $N \to \infty$. The common bias term, that is, $O(h_{1}^{2} + h_{2}^{2} + h_{3}^{2})$,  vanishes under the condition. However, our main conclusion still holds if the assumption is relaxed.  
Importantly, the asymptotic variance $\V{x^l}$ can be estimated using the plug-in method given by. 
% Specifically, % when $W$ is continuous,
%the estimated asymptotic variance for $\hat \tau(x^{l})$ is
{\small 
 \begin{equation} \label{asy-var}  
\hat{\mathbb{V}}(x^l) := \frac{1}{Nh_3 \hat{f}(x^l) } \paren{  \nu \cdot \hat{\mathbb{V}} \paren{ \beta(X^{l}, e) | X^l = x^l}  +   \int \bar{K}^{2}(t)dt \cdot  \hat{\mathbb{E}} \paren{ \frac{ (D - e)^{2}}{e^{2} (1 - e)^{2}} \xi^{2} | X^l = x^l  } },      
 \end{equation}}
where $\hat f(x^l)$ is the kernel density estimation, $\hat{\mathbb{V}}(\beta_{1}(X^{l}, e)| X^{l} = x^{l})$ can be estimated by conducting a nonparametric regression of $(\hat \beta(X^{l}, e) - \hat \tau(X^{l}))^{2}$  on $X^{l}$ \citep{Fan-Yao-1998}. Because $\xi$ is estimated using the residual in model (\ref{beta working}),  we obtain $\hat  E\big [
\xi^{2} (D - e)^{2} \big /e^{2} (1 - e)^{2}  | X^{l} = x^{l} \big ]$  by regressing $ (D -\hat e)^{2} \hat \xi^{2} \big / \hat e^2 (1 - \hat e)^2$  on
$X^{l}$.  In addition, $\nu$ and $\int \bar{K}^{2}(x)dx $ can be calculated directly. 
% For example, for Gaussian kernel, $ \nu = \int K^{2}(t)dt  =1/\sqrt{4\pi}$.
% and equation \bcol{(\ref{kernel_bar})} presents the closed form of
%$\bar{K}(x)$ and $\int \bar{K}^{2}(x)dx$. 
Finally, we have the following conclusion.
\begin{corollary} The estimated asymptotic variance is a consistent estimator, namely,    
			$ \hat{\mathbb{V}}(x^l)  - \mathbb{V}(x^l) = o_{p}(1).    $
\end{corollary}

% \pw{The preceding  estimator and theoretical analysis  focus on the case of continuous $X^{l}$. Actually,  the proposed estimation procedure in Section 3.1 accommodate to different data types of $X^{l}$ and $Y$.  We extend to the PSR method to the case of discrete $X^{l}$ by adopting kernel smoothing method for discrete variable \citep{Aitchison-Aitken-1976, Li-Racine-2010}, corresponding estimator, asymptotic properties and simulation study are presented in Section 7 of the Supplementary Material.}

% ===============================================================
\section{Simulation studies}  \label{sec: sim} 
% nonparametric 
We conduct extensive simulation studies to assess the finite-sample performance of the PSR, compared with the existing IPW method of \cite{Abrevaya-Hsu-Lieli-2015}, and the AIPW method of \citet{Lee-Okui-Whang-2017}. In addition, we consider a matching variant of the PSR, where we use a matching method to estimate $\beta(X^{l}, e)$. Specifically, we replace Step 1 in the PSR by first creating matched pairs using matching on $(X^{l}, \hat e(X))$, and then using the matched pairs to impute the missing potential outcomes. Finally, we calculate $\beta(X^{l}, e)$ using the imputed potential outcomes. In the matching, we use one-to-one matching and the Mahalanobis metric.  
In the following, we focus on settings with extreme propensity score values. We also consider an alternative scenario in which propensity scores are distributed far from zero and one. 

 % newly proposed AIPW approaches of \citet{Lee-Okui-Whang-2017}, \cite{Fan-Hsu-Lieli-Zhang-2019} and \cite{Zimmert-Lechner-2019}.  
 \subsection{Simulation Setup} 
We set $l = 1$ and the covariate dimension $p=$ 5, 20, and 50. $X^{l} \sim$ Uniform$(-0.5, 0.5)$ and $X^{-l} = (X^{-l}_{1}, ..., X^{-l}_{p-1}) \sim$ Norm$(0, \Sigma)$, with $\Sigma_{j, k} = 2^{-|j - k|}$ for $1 \leq j, k \leq p-1$. The assignment of treatment $D$ follows the logistic model, with $\Prob{D =1 | X}  =\exp( X^{\intercal} \alpha) / (1+\exp( X^{\intercal} \alpha))$. 
 
 In the setting with extreme values of propensity scores, we consider two assignment mechanisms: Mechanism A, $\alpha = (1,-1,-1,1,-1, 0, \ldots,0)^{\intercal}$, with five nonzero entries; Mechanism B, $\alpha= (1, 1, 1,1, 1,0, \ldots,0)^{\intercal}$, with five nonzero entries.       
We conduct four simulation settings, in which the heterogeneous treatment effects $\tau(x^{l})$ are linear, quadratic, polynomial,  and complex functions of $x^l$, respectively.  The potential outcomes under the four contexts are modeled as follows:  

 \begin{enumerate}[label= \Roman*:]  
 \item   $Y(1) =  X^{l} (1 + 2 X^{l})^{2} (X^{l} - 1)^{2} + f(X) + \epsilon(1)$, $Y(0) = f(X) + \epsilon(0)$, 
 $f(X) = (X^{l})^{2} X_{1}^{-l} X_{2}^{-l} X_{3}^{-l} X_{4}^{-l}$,   
 
\item  $Y(1) =  X^{l} (1- X^{l}) \cos(X^{l}) \log(X^{l} + 2) \exp(X^{l}) + f(X) + \epsilon(1)$, 
$Y(0) = f(X) + \epsilon(0)$,  $f(X) = (X^{l})^{2} X_{1}^{-l} X_{2}^{-l} X_{3}^{-l} X_{4}^{-l}$.

\item  $Y(1) = X^{l} + f(X) + \epsilon(1)$, $Y(0) = f(X) + \epsilon(0)$, $f(X) =\{ X^{l} X^{-l}_{1} + \exp( X^{-l}_{2} - 3 ) ( \sin(X^{-l}_{3} ) + \cos(X^{-l}_{4}) )\} /2$,  
 
\item  $Y(1) =  5 (X^{l})^{2} + X^{l} + f(X) + \epsilon(1)$, $Y(0) = f(X)+ \epsilon(0)$, $f(X) = (X^{l})^{2} \{ \sum_{j=1}^{4} X^{-l}_{j} / 2^{j+1} \} $.

\end{enumerate}
The error terms $\epsilon(1)$  and  $\epsilon(0)$ are independently and identically distributed (i.i.d.) with Norm$(0, 1)$. The true $\tau(x^{l})$ in the four simulation settings are $x^{l} (2 x^{l} + 1)^{2} (x^{l} - 1)^{2}$,  $ x^{l} (1- x^{l}) \cos(x^{l}) \log(x^{l} + 2) \exp(x^{l})$, $x^{l}$, and $5 (x^{l})^{2} + x^{l}$, respectively. Note that the potential outcome models have a complex form, making it difficult to correctly specify the outcome models. 

\begin{figure}[h!]
   \centering
    \includegraphics[scale = 0.55]{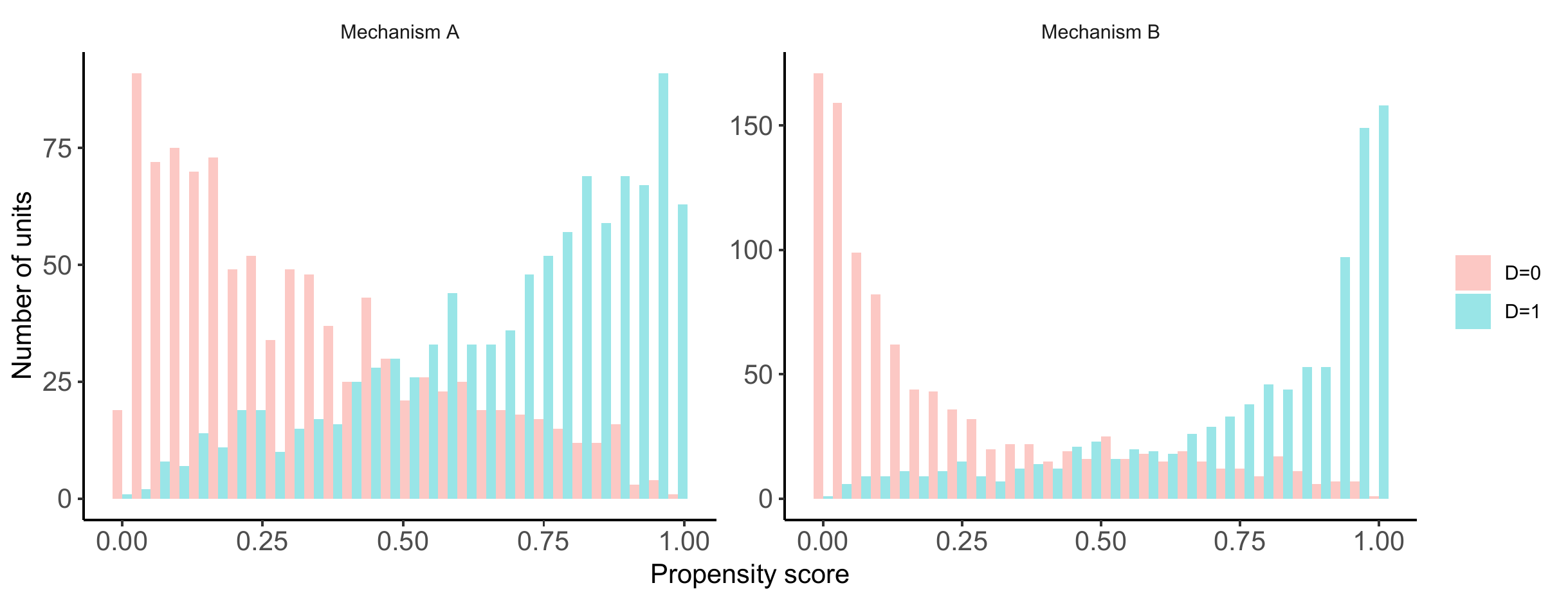}
        \vskip -0.2cm
    \caption{Distribution of the true propensity score under the two treatment assignment mechanisms ($N = 2000, p = 5$).}
      \label{ps-distri}
\end{figure}

Simulations I and II use assignment Mechanism A, and  Simulations III and IV use assignment Mechanism B. The distributions of the propensity scores under the two mechanisms are plotted in Figure  \ref{ps-distri}. We observe that many propensity score values are close to zero or one, making the weighting-based propensity score methods highly unstable in the context. 
%\pw{These scenarios can cause a lot of instability and challenges when using the inverse of propensity score as a weight.  In addition, there are more values of propensity score close to 0 or 1 in Mechanism A than those in Mechanism B.}

We estimate the propensity scores using logistic regression. The bandwidths are chosen using existing methods \citep{Li-Racine-2007, Ruppert-etal-1995}, and implemented using the \texttt{R} functions \texttt{npscoefbw}  from the package  \textbf{np} \citep{Racine2021} and \texttt{dpill}  from the package  \textbf{KernSmooth} \citep{Wand-etal2021}.  For the competing AIPW method,  the outcome regression functions are estimated using linear models.   
 Each simulation is based on 1000 replicates, with sample sizes 500, 1000, and 2000.   
 
 	We evaluate the performance of the PSR using the sample average bias (Bias), sample average standard deviation (SD),  mean absolute error (MAE), mean squared error (MSE), and average 95\% confidence interval coverage proportion (CP95). For the PSR, CP95 is estimated using the asymptotic variance formula (\ref{asy-var}). For the IPW and AIPW methods, CP95 is estimated as in \cite{Abrevaya-Hsu-Lieli-2015} and \citet{Lee-Okui-Whang-2017}, respectively. Finally, for the matching variant of the PSR and random forest methods, CP95 is estimated using 100 bootstraps. 
 % and \textcolor{blue}{compare it with the other two approaches with  based on 1000 simulations}. %\rcol{The CP95 is calculated by }. \citep{Racine2021} and \citep{Wand-etal2021}
%sample average bias
%for the nonparametric function $\tau(x^{l})$ based on 100 equally spaced grid points on the 
% range of $[-0.5, 0.5]$, 

Table \ref{tab1} summarizes the results of the PSR for cases I--IV. We observe that as the sample size increases,  MAE and MSE decrease, and CP95 becomes closer to the nominal value of 0.95.   
Moreover, the results are similar for different values of $p$, suggesting that the PSR is insensitive to the dimension of the covariates.

\begin{table}
\caption{\label{tab1} The performance of the PSR for cases I--IV.}
\centering
 \renewcommand\arraystretch{1}
 \scriptsize
 %\tiny 
%\setlength\tabcolsep{3pt}
 \fbox{ 
 \begin{tabular}{c  ccccc   ccccc  ccccc } 
  %&  \multicolumn{12}{c}{ }    \\     \hline
   $(N, p)$    & Bias (SD) & MAE & MSE  & CP95 &   & Bias (SD) & MAE & MSE & CP95  \\
   &  $\times10^{-2}$ & $\times10^{-2}$&$\times10^{-2}$  &\% &   &$\times10^{-2}$ & $\times10^{-2}$ & $\times10^{-2}$  &\% \\
        \hline  
         &   \multicolumn{4}{c}{Simulation I}  & &  \multicolumn{4}{c}{Simulation II}    \\
  $(500, 5)$ & 0.0 (16.9)  & 13.2 & 2.8 & 87.4 & & 0.1 (15.5)  & 12.3 & 2.4 & 86.5  \\
    $(1000, 5)$  & -0.4 (12.7)  &  10.0 & 1.6 & 91.8 & & 0.1 (11.8)  &  9.3 & 1.4 & 90.7  \\
 $(2000, 5)$ &  -0.5 (9.4)  &  7.5 & 0.9 & 94.4  &   & 0.2 (8.4) & 6.6 & 0.7 & 93.6 \\
  $(500, 20)$ & 0.0 (17.3)  &  13.7 & 3.0 & 87.5 &  & 0.4 (15.4) & 12.4 & 2.6 & 86.5     \\
    $(1000, 20)$  & -0.0 (12.7)  &10.3 & 1.7 & 89.8  &  & -0.0 (11.6) & 9.2 & 1.4 & 90.6  \\
 $(2000, 20)$  & 0.0 (9.6)  & 7.5 & 0.9 & 94.3 &   & -0.0 (9.0)  & 6.8 & 0.7 & 93.8   \\
  $(500, 50)$ &  -0.4 (17.3)  &  13.7 & 3.1 & 87.3 & & -0.0 (16.5)  & 12.7 & 2.6 & 86.9  \\
    $(1000, 50)$  & -0.2 (13.0) & 10.2 & 1.7 & 91.5 & & 0.0 (11.8)  & 9.6 & 1.5 & 90.4  \\
 $(2000, 50)$   & -0.3 (9.4) & 7.5 & 0.9 & 94.2 &  & 0.0 (9.1) & 6.8 & 0.8 & 93.8  \\
    %    \hline
            \hline  
             &  \multicolumn{4}{c}{Simulation III} &&  \multicolumn{4}{c}{Simulation IV}  \\
  $(500, 5)$ &   3.3 (17.4)  & 13.7 & 3.1 & 88.1 &  & 2.0 (22.1)  & 17.2 & 4.9 & 88.9 \\
    $(1000, 5)$  &  2.7 (13.2)  & 10.6 & 1.8 & 91.4 &  & 1.4 (16.8)  & 13.2 & 2.8 & 92.2 \\
 $(2000, 5)$ &  3.2 (10.5)  & 8.6 & 1.2 & 93.9 &   & 1.6 (13.0) & 10.2 & 1.7 & 95.2 \\
  $(500, 20)$ &  2.8 (17.8) & 14.2 & 3.4 & 86.9 & & 2.1 (22.3) & 17.5 & 5.1 & 88.6 \\
    $(1000, 20)$  &  2.8 (13.9) & 10.8 & 1.9 & 91.0 &  & 1.6 (16.9) & 13.2 & 2.9 & 92.7\\
 $(2000, 20)$  &  2.5 (10.7) & 8.3 & 1.1 & 93.4 & &  1.5 (12.8) & 10.1 & 1.7 & 95.1 \\
  $(500, 50)$ &   2.9 (19.4) & 14.6 & 3.5 & 85.8 & & 1.8 (22.1)   & 17.4 & 4.9 & 88.3 \\
    $(1000, 50)$  &  3.1 (14.4) & 11.4 & 2.1 & 90.2 & & 1.2 (16.7)  & 13.0 & 2.8 & 92.2   \\
 $(2000, 50)$    & 2.4 (10.6) & 8.6 & 1.2 & 93.7 & & 1.3 (13.2)  & 10.4 & 1.8 & 94.6 \\
    \end{tabular} }
%   \begin{flushleft}
%     Note: \rcol{CP95 represents  the average 95\% confidence interval coverage proportion, estimated using the asymptotic variance formula (\ref{asy-var})}.
%   \end{flushleft} 
\end{table}

\begin{figure}[h!]
   \centering
    \includegraphics[scale = 0.55]{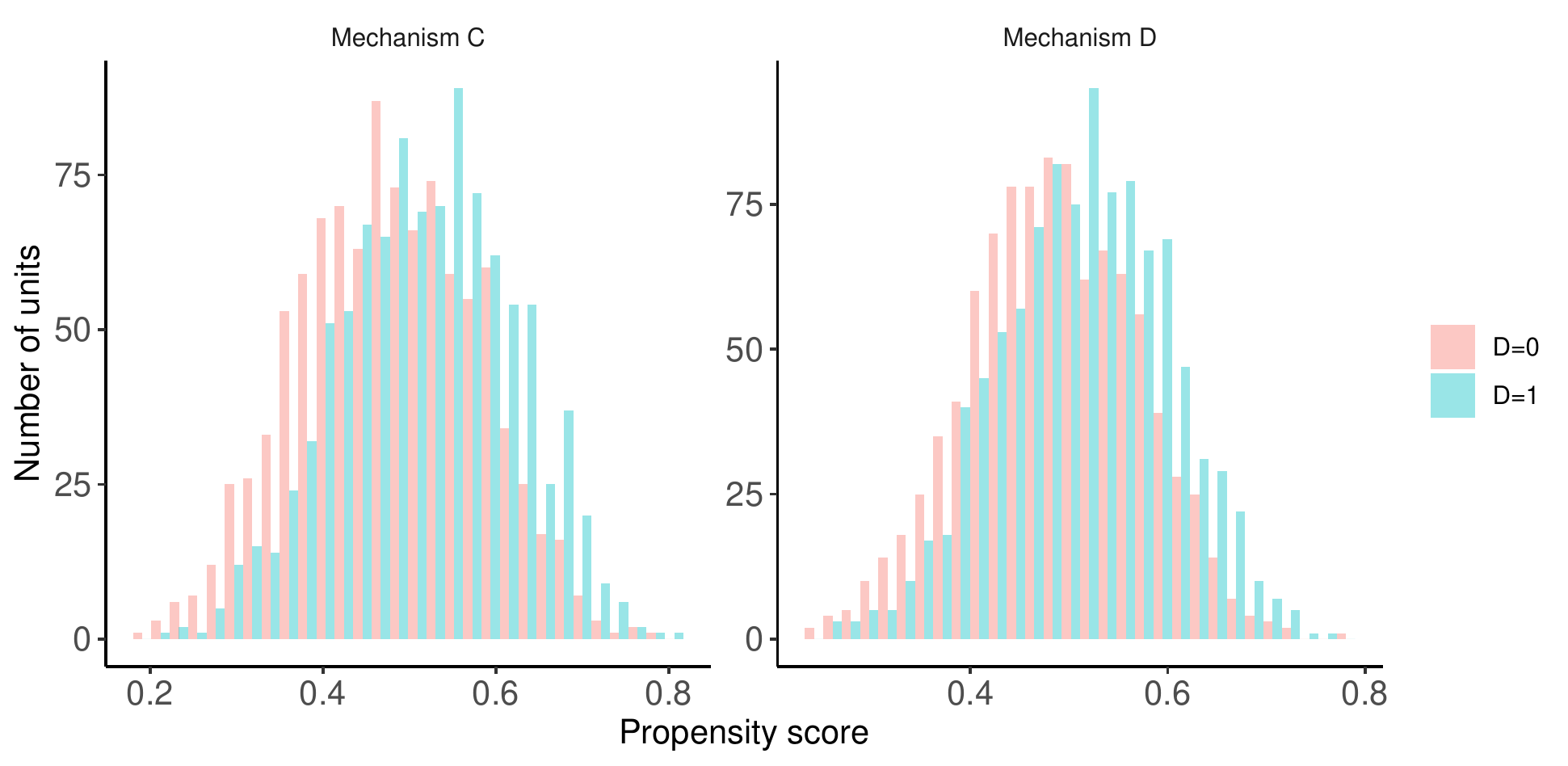}
        \vskip -0.2cm
    \caption{Distribution of the true propensity score under the treatment assignment mechanisms C and D ($N = 2000, p = 5$).}
      \label{ps-distri2}
\end{figure}

We also consider an alternative scenario in which the propensity scores are distributed far from zero and one. We replace the data-generation mechanisms in Simulations I--IV with new mechanisms. Specifically, we replace Mechanism A in Simulations I and II with Mechanism C, where $\alpha$ is set as $0.25 (1,-1,-1,1,-1, 0, ..., 0)^{\intercal}$ with five nonzero entries,  and denote them as Simulations V and VI, respectively. Then, we replace Mechanism B in Simulations III and IV with Mechanism D, $\alpha = 0.125 (1,-1,-1,1,-1, 0, \ldots,0)^{\intercal}$ with five nonzero entries, and denote them as Simulations VII and VIII, respectively. Figure \ref{ps-distri2} shows the propensity score distributions of the treatment assignment mechanisms C and D.

 % Figure \ref{ps-distri2} shows the propensity score distributions of the treatment assignment mechanisms C and D.    

\begin{table}
\caption{\label{tab4} The performance of the PSR, IPW,  AIPW, and matching variant of the PSR under different simulation settings ($N=2000,  p =5$)}.
\centering
 \renewcommand\arraystretch{1}
\scriptsize
 \fbox{
 \resizebox{\linewidth}{!}{\begin{tabular}{c c  ccccc   ccccc} 
Pair &  Case  & Bias (SD) &  MAE & MSE  & CP95 &  & Case  & Bias (SD) &  MAE & MSE  & CP95  \\
&  & $\times10^{-2}$ & $\times10^{-2}$&$\times10^{-2}$  &\% &  &   &$\times10^{-2}$ & $\times10^{-2}$ & $\times10^{-2}$  &\% \\
        \hline 
             & &   \multicolumn{10}{c}{PSR method}  \\   
(1)  & I &   -0.5 (9.4)  &  7.5 & 0.9 & 94.4  &   &  V & -0.2 (8.9) &  6.9 & 0.8 & 93.7    \\  
(2)  &  II &  0.2 (8.4)  & 6.6 & 0.7 & 93.6  & & VI &   0.0 (7.7) & 6.1  & 0.6  & 94.0    \\
(3)  &  III & 3.2 (10.5)   &  8.6 & 1.2 & 93.9  & & VII  &  1.1 (8.3) & 6.7 &  0.7 & 93.2      \\
(4)  & IV &  1.6 (13.0)  & 10.2 & 1.7 & 95.2   && VIII  & 1.1 (12.1) & 9.6 & 1.5 & 94.1   \\   
\hline  
      & &   \multicolumn{10}{c}{IPW method}  \\  
(1)  & I &   0.6 (33.9)  &  14.7 & 11.5  & 92.7 &   &  V & 0.6 (11.5)   &  8.6  & 1.3  &  91.1     \\  
(2)  & II &  0.3 (31.1) &  14.2 & 9.7 &  93.7  & &  VI & 0.0 (10.9)    & 8.1   & 1.2   &  91.8   \\
(3)  &  III &   -0.1 (53.8) &  20.7 & 28.9 &  94.2  & & VII  & -0.2 (10.7)   & 8.1  &  1.1  &  92.1      \\
(4)  &IV &  2.2 (49.0)  & 22.3 & 24.1 &  93.9   &&VIII  & 2.4 (11.7)  & 9.1  & 1.4  &  91.2 \\    
       & &   \multicolumn{10}{c}{AIPW method}  \\  
(1)  & I &  0.4 (29.8)   &    17.0  &  8.9 &  94.9  &  & V & -0.2 (13.7)  & 10.4  & 1.9 &  94.0     \\  
(2)  & II & -0.2 (26.9)   &  17.0 & 7.2 &  94.6  & & VI & -0.0 (13.9)   & 10.4    & 1.9   &  94.3     \\
(3)  & III &   0.8 (66.7)  &  25.6  & 44.6  &  94.0  & &  VII  & -0.1 (13.3)  & 10.1  & 1.8  &  93.8    \\
(4)  & IV &  0.5 (61.8)  & 25.9  & 38.2  &  95.0  & &  VIII  & 0.6 (14.0)  &  10.8 & 2.0 &  94.1   \\
       & &   \multicolumn{10}{c}{Matching variant of PSR}  \\  
(1)  & I &   0.4 (22.7) &   17.7 & 5.2  &  93.4      &  & V &  0.0 (18.0) &  13.9 & 3.2 &   96.1     \\  
(2)  & II &  0.5 (22.8) & 17.8 & 5.2   &  93.2     & & VI &  0.3 (17.9) & 13.8 & 3.2  & 96.2     \\
(3)  & III &  0.7 (29.3) & 23.1 & 8.6 &  90.4       & &  VII  &  0.2 (18.0) & 14.0 & 3.2  & 95.8      \\
(4)  & IV &   0.7 (29.2) & 23.0 & 8.5 &  90.5      & &  VIII  & 0.7 (17.8) & 13.9 & 3.2  & 95.8   \\      
    \end{tabular}}}  
%       \begin{flushleft}
%     Note: All the values in this table have been magnified 100 times. 
%   \end{flushleft} 
     \end{table}

We present the results for Simulations V--VIII, and contrast them with those of Simulations I--IV in Table \ref{tab4} ($N = 2000, p = 5$). Each row represents the results for a pair of simulations that differ only in terms of their propensity score mechanisms. We find no significant differences between Bias (SD), MAE, and MSE when PSR is used.  However, when we use the IPW, AIPW, and the matching variant of the PSR, the SD, MAE, and MSE in settings with extreme propensity score values are significantly different from those in settings with general propensity scores. This again shows that the PSR is robust to extreme propensity score values. In addition, 
we find that the matching variant of the PSR performs similarly to our PSR in terms of bias, but has a larger SD, MAE, and MSE,   because both methods leverage the idea of propensity score matching. However,  unlike the matching method, the PSR uses propensity scores in a nonparametric manner and smoothly estimates $ \beta(x^{l}, e)$, making it less sensitive to minor differences between inexact matches. This explains why the PSR has a smaller SD, MAE, and MSE than those of its matching variant.

%\rcol{Compared to the other two approaches, PSR has mean errors closer to 0 with smaller variances}. This phenomenon is observed in all four simulation settings, indicating that PSR is indeed robust to extreme values of propensity scores.
  
%Table \ref{tab2}  compares the MAE and MSE of the three approaches: PSR, IPW and AIPW.  \rcol{Compared to the other two approaches, PSR has mean errors closer to 0 with smaller variances}. This phenomenon is observed in all four simulation settings, indicating that PSR is indeed robust to extreme values of propensity scores.

% ===========================================

\subsection{Alternative estimation methods on propensity scores}

% We have added simulation analyses considering two types of estimation errors on propensity scores.  
% \textcolor{blue}{We also conduct additional simulation to assess the robustness of the proposed estimators under different estimation errors of propensity score.}  
 We consider two alternative scenarios when estimating the propensity score. In the first scenario, we estimate the propensity scores using a probit model. In the second scenario, we estimate the propensity scores nonparametrically, using the random forest method (R package \textbf{grf}). We compare the two with the baseline scenario in which the propensity scores are estimated using the true logit model. Our results show that  the obtained Bias, SD, MAE, and MSE are all very close under the three scenarios. % \rcol{However, as expected, the CP95 calculated by (3.13) under the random forest performs is lower than 95$\%$, this is because the Assumption I may not be satisifed?(TODO).}  We show this in Table \ref{tab3}.

\begin{table}
\caption{\label{tab3} The performance of the PSR under different estimation errors of propensity scores ($N=2000,  p =5$).}
\centering
 \renewcommand\arraystretch{1}
 \scriptsize
 %\tiny 
%\setlength\tabcolsep{3pt}
% \scriptsize 
 \fbox{
 \resizebox{\linewidth}{!}{\begin{tabular}{c ccccc   ccccc ccccc  cccc} 
%  \hline
Method   & Bias (SD) &  MAE & MSE  & CP95 &    & Bias (SD) &  MAE & MSE & CP95        \\
   & $\times10^{-2} $&  $\times10^{-2}$ & $\times10^{-2}$  &\% &    & $\times10^{-2}$ &  $\times10^{-2}$ &$\times10^{-2}$ & \%    \\
  \hline
     &   \multicolumn{4}{c}{Simulation I}  & &    \multicolumn{4}{c}{Simulation II}    \\  
Logistic  &   -0.5 (9.4) &  7.5 & 0.9 & 94.4 & & 0.2 (8.4)  & 6.6 & 0.7 & 93.6   \\
 Probit  &  -0.1 (9.6) & 7.5 & 0.9 & 94.1 & &     0.1 (8.9) & 6.9 & 0.8 & 93.9  \\ 
 % Random Forest & 0.2 (9.6) & 7.6 & 0.9 & 86.9 &&  0.0 (8.7) & 6.9 & 0.8  & 85.8 & &  3.0 (10.6) & 8.8 & 1.2 & 85.9 & &  2.5 (12.2) & 10.0 & 1.6 & 91.6   \\ 
  Random Forest  & 0.2 (9.6) & 7.6 & 0.9 & 91.4 &&  0.0 (8.7) & 6.9 & 0.8  & 92.7  \\ 
 %  \hline 
   \hline
      &   \multicolumn{4}{c}{Simulation III}  &  &  \multicolumn{4}{c}{Simulation IV}  \\  
Logistic  &     3.2 (10.5)  &  8.6   & 1.2 & 93.9 &  & 1.6 (13.0) & 10.2 & 1.7 & 95.2  \\
 Probit  & 3.2 (10.6) & 8.7 & 1.2 & 93.8  & &  1.7 (13.1) & 10.3 & 1.7 & 94.6   \\ 
 % Random Forest & 0.2 (9.6) & 7.6 & 0.9 & 86.9 &&  0.0 (8.7) & 6.9 & 0.8  & 85.8 & &  3.0 (10.6) & 8.8 & 1.2 & 85.9 & &  2.5 (12.2) & 10.0 & 1.6 & 91.6   \\ 
  Random Forest  &   3.0 (10.6) & 8.8 & 1.2 & 91.2 & &  2.5 (12.2) & 10.0 & 1.6 & 91.6   
\end{tabular}}}
%   \begin{flushleft}
%     Note:\rcol{CP95 represents the average 95\% confidence interval coverage proportion. CP95 for Logistic, Probit and Random Forest was estimated using the asymptotic variance formula (\ref{asy-var}) and for Random Forest II was estimated using the Bootstrap. }. % \bcol{The symbol ``---'' means non-computability.} 
%   \end{flushleft} 
\end{table}

% ================================================================
\section{Application}\label{real example}

We demonstrate our method in two studies using data from the National 2009 H1N1 Flu Survey (NHFS). The NHFS was a large one-time telephone survey conducted in the United States from October 2009 through June 2010, by the Centers for Disease Control and Prevention (CDC). The survey asked questions on participants' seasonal influenza vaccination status, whether had been sick with an influenza-like illness in the past month, the number of days they had taken off work owing to influenza, whether they have paid sick leave benefits, the number of times they see a doctor, as well as other relevant information (e.g., influenza-related behaviors, opinions about influenza vaccine safety and effectiveness, the size of the household, and demographic characteristics) \citep{Questionnaire2010}. The NHFS public dataset has been released by the CDC, National Center for Immunization and Respiratory Diseases (NCRID), and National Center for Health Statistics (NCHS). The datasets are used to analyze the vaccination coverage, vaccination beliefs, and behaviors \citep{CDC2010, Ding-Santibanez2011, Burger-Reither2021}. Using a subset of the data comprising adults (i.e., age $\geq$ 18) and English-speaking participants, we conducted the following two studies. % as detailed in the following.

% \pw{Underline the strengths of our proposed methods again....}
%I in the second study, we assess the impact of having paid sick leave on the number of times visiting doctors.

\subsection{Effect of seasonal influenza vaccination on the number of sick days}
We estimate the effect of seasonal influenza vaccinations on the number of sick days taken because of an influenza-like illness. Our primary outcome is the number of days taken off work (after taking ``log'') when sick with an influenza-like illness, as reported by the participants during the interview. We consider a subset of the data, including only adult participants who had reported being infected with an influenza-like illness,  and excluding participants with missing outcomes or treatments. We consider covariates that have non-missing values for at least 70\% of the participants.  For each selected covariate, the missing value is treated as a new category.
% 	% For some missing covariates,  % we treat the missing values  some of them are  
 % we treated the missing values as a new category and incorporated them into the model, as the missing values provide information that  may be helpful to fit propensity score model, these variables include 
% \texttt{VACC\_PNEU\_COUNT},  \texttt{INT\_H1N1\_DKNW\_F},\texttt{INT\_H1N1\_DNOT\_F},\texttt{INT\_H1N1\_DYES\_F},\texttt{INT\_H1N1\_PNOT\_F},\texttt{INT\_H1N1\_PYES\_F},\texttt{INT\_H1N1\_REFD\_F},\texttt{DOCREC\_BOTH\_F},\texttt{DOCREC\_DKNW\_F}, \texttt{DOCREC\_H1N1\_F},\texttt{DOCREC\_NTHR\_F},\texttt{DOCREC\_REFD\_F},\texttt{DOCREC\_SEAS\_F},\texttt{PATIENT\_CONTACT\_F},\texttt{INC\_CAT1}. 	   	
%	and then removed the individuals with missing covariates.  
%	 Furthermore, we eliminated invalid covariates, such as person identifier, household identifier,  and categorical variables with only a single class, and some extremely unbalanced binary variables (1 v.s. 2441).   
Our final sample comprised 2442 participants and 65 (i.e., $p = 65$) dimensional covariates, where 1145 individuals have had a seasonal flu vaccination.  We include all the informative covariates in the analysis, because conditioning on any given covariates is, in general, better than not conditioning \citep{Rosenbaum2002, Rubin2009, Ding-Miratrix2015}. Nevertheless, the potential bias introduced by the adjustment needs further attention \citep{Pearl2015, Ding2015}. Descriptive statistics of the sample and the covariates are listed in the Supplementary Material Table S2. % , 
% (\bcol{COMMENT: For the supplementary materials, (1) add one note-row under Table S2. Add the sentence``Notes: Variable names are given in \cite{Questionnaire2010}." (3) Table 2 caption – state N \& p.(3) Correct ``Table S3" into ``Table S2 (Continue)". })
% in response to the question, ``About how many days did you miss school or work because of this (influenza-like-illness) illness?'' 

\begin{figure}
\centering    
\makebox{\includegraphics[scale = 0.55]{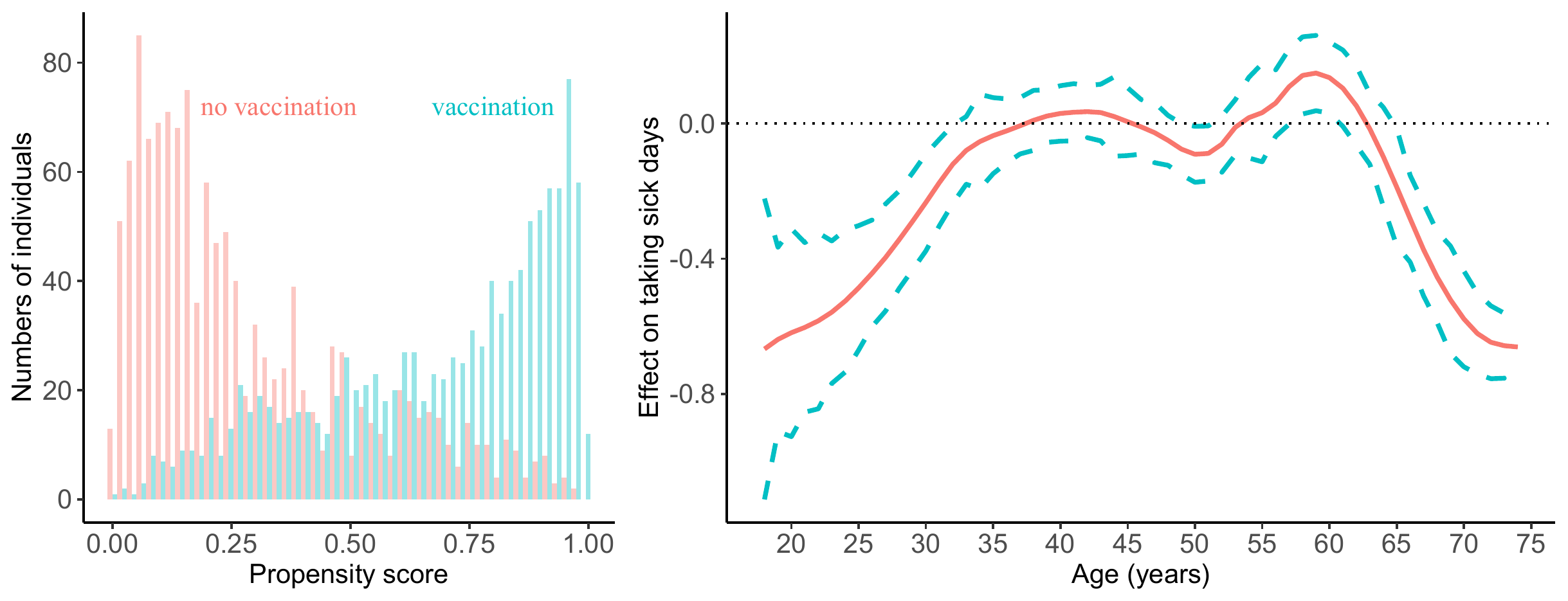}}
\caption{\label{fig3} (a) Distribution of the propensity scores. (b) Effects of seasonal vaccination on taking sick days across age groups. Dashed lines refer to 95\% confidence interval.}
\end{figure}

 For our approach, we follow the three steps described in Section 3.1. We estimate the propensity scores using a logistic regression and the heterogeneous effects using a standard local linear regression, as described in Section 3.2. We display the results in Figure \ref{fig3}.
 
%\begin{table}
% \caption{\label{tab3} Sample quantiles of  estimated propensity score values.}
%\centering
%\scriptsize
%\fbox{%
%\begin{tabular}{ l ccccc }  \hline 
%	group &  minimum  &  $25\%$Q    &  median   &  $75\%$Q     &    maximum     \\
%			 \hline
%   $D=1$   & 0.000035 & 0.212288 & 0.485091 & 0.799593 & 0.999947 \\ 
%   $D = 0$  & 0.001843 & 0.417812 & 0.504459 & 0.601570 & 0.998410 \\ 
%	 \hline
%\end{tabular}
%}
%\end{table}
 
 %\tabnote{Note: 25\%Q and 75\%Q denote  25\% and 75\% sample quantiles.}
%\label{table1}

 As shown in Figure \ref{fig3}(a),  the propensity score values vary from 0.001 to $0.994$ for the group of people with a seasonal vaccination, and from 0.001 to $0.978$ for the group of people without a seasonal vaccination. Therefore, we cannot use methods that are sensitive to propensity scores. Using our PSR method, we show that the effects of seasonal vaccination on taking sick days vary across age groups (Figure \ref{fig3}b). Furthermore, seasonal vaccines decrease the number of sick days off work for the population aged over 60 and under 35,  but have a negligible impact on the other age groups. Because seasonal vaccines prevent severe symptoms of influenza \citep{Deiss-etal2015}, in general, people with seasonal vaccinations are more likely to develop light symptoms and, therefore, are less likely to take leave off work when sick with influenza. However, our results suggest that people aged 36--59 are equally likely to leave work to see doctors, regardless of the severity of the symptoms.

\subsection{Effect of having paid sick leave on visiting doctors}

\begin{figure}
\centering   
\makebox{\includegraphics[scale = 0.55]{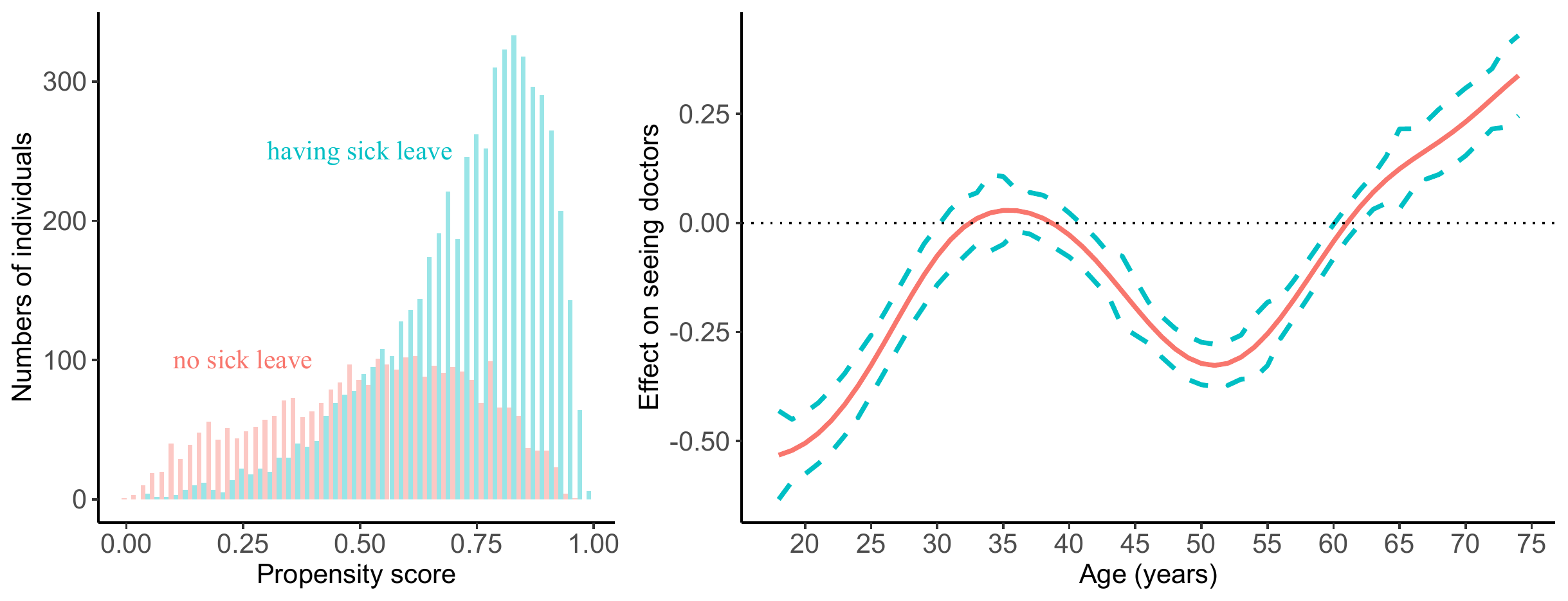}}
\caption{\label{fig4} (a) Distribution of the propensity scores. (b) Effects of having paid sick leave on seeing a doctor, across age groups. Dashed lines refer to the 95\% confidence interval.}
\end{figure}

In the second study, we estimate the effect of having paid sick leave on the number of times people see a doctor regardless of the disease type. Our treatment is whether the adult earns paid sick time off from employment, with $D=1$ representing having paid sick leave, and $D=0$ representing not having paid sick leave. The primary outcome $Y$ is the self-reported number of times a person sees a doctor. During the interview, the participants were asked to provide the number of times they had seen a doctor or other health professional about health since August 2009. For the study, we also consider a subset of the data, including only adults whose paid sick leave indicator $D$ is known. Our final sample comprised 8425 participants and 62 (i.e., $p = 62$) dimensional covariates, where 5502 have paid sick leave and 2923 do not. Descriptive statistics of the sample and covariates are listed in the Supplementary Table  
  S3. 
%in response to the question, ``About how many days did you miss school or work because of this (influenza-like-illness) illness?'' 
 Similarly, we estimate the propensity scores using the logistic models and the heterogeneous effects using the standard local linear regression, as described in Section 3.2. We display the results in Fig. \ref{fig4}. 

As in the first study, the propensity scores are distributed with values varying from 0.033 to 0.984 for the group of people with paid sick leave, and 0.001 to $0.956$ for people without paid sick leave. %( Fig. \ref{fig: illustration 2 propensity scores}).
  The effects of having paid sick leave vary substantially across age groups (Fig. \ref{fig4}b). Specifically, having paid sick leave increases the number of times of seeing doctors for people aged over 65, but has no significant effects for people aged 33--40. Interestingly, having paid sick leave motivates people aged under 33 or aged 41--64 to reduce the number of times they see doctors. One possible explanation is that these people may have a bundle of paid time-off benefits that combines sick days, vacation days, and other types of leave, where reducing the number of times they see a doctor may increase their overall benefits \citep{Zhai-etal2018, Smith-Kim2010}.

\bigskip  \bigskip

\section{Discussion} \label{sec: discussion} 
% \pw{Propensity scores can be estimated nonparametrically}, through e.g., the kernel regression \citep{Zhou-Elliott-Little-2019}.    
We have proposed a nonparametric PSR method for estimating the heterogeneous treatment effects in a wide context, including settings in which the propensity scores are close to zero or one and the number of full covariates is large. We have established the large-sample properties, and show that it outperforms existing methods in our simulation studies. Although in the main text we consider continuous $X^{l}$, our methods hold regardless of the type of variable $X^{l}$. In \S 7 of the Supplementary Material, we present theoretical results also for discrete $X^{l}$, where we use the typical kernel smoothing method for the estimation \citep{Aitchison-Aitken-1976, Li-Racine-2010}.

 Note that our theoretical results hold when we replace the parametric model specification of the propensity score with a semiparametric model, such as the single index model. In this case, we may simply use $X^{\intercal} \hat \alpha$ instead of $e(X)$, because the PSR is built on the balancing property of the propensity score, and any one-to-one function of propensity scores has the same balancing property. For a single index model, many available estimators satisfy the 
assumptions in Assumption 1 with \emph{unknown} link functions;  see, for example, \cite{Horowitz-Hardle1996}, \cite{Ichimur1993}, \cite{Klein-Spady1993}, \cite{Hardle-Spokoiny-Sperlich1997}, and \cite{Wang-Yang2009} for further detail.  Furthermore, methods that can improve the balance property  \citep{Huang-Chan2017, Wei-etal2017, Tan2020BKA,Imai-Ratkovic2014, Ning-etal2020} could be useful for improving the performance of the PSR. 
% Carroll-Fan-etal1997, Hardle-etal1993, Hardle-Stoker1989,  Xia-Li1999  Powell-etal1989,

% \bcol{We emphasize that our methods do not rely on the correct model specification of propensity scores. The simulation studies have shown that misspecification of the propensity score model has negligible impacts. 
% Moreover, the propensity score can be estimated nonparametrically. As the earlier results by \cite{Mammen-etal2012} have shown, $\hat \tau(x^{l})$ could be consistent as long as the estimator of propensity score is uniformly consistent --- a weaker condition than the $\sqrt{N}$-a consistent condition in Assumption 1(i)}. 
 
%However, we note that in real context PSR relies on estimated propensity scores, and \rcol{having a consistent estimation of propensity scores is crucial to ensuring asymptomatic consistency.  }We have provided illustrations where propensity scores are estimated nonparametrically. 
%Bandwidth selection is an important component in nonparametric kernel-based methods. 

For nonparametric estimation, we use kernel-based methods, for which an appropriate choice of the bandwidths,  $h_{1}$, $h_{2}$, and $h_{3}$, is important to achieve good accuracy. In the simulation, we simply use the existing bandwidth-selection methods \citep{Li-Racine-2007}, and the bandwidths are chosen independently in the corresponding varying coefficient models or local linear models. To improve accuracy, approaches that simultaneously account for the bandwidth choice in estimating $\beta(X^{l}, e)$ and $\tau(X^{l})$ are probably helpful.  % Ruppert-etal-1995

%  variables
Finally, we have focused on continuous outcomes. However, the PSR is not restricted to such outcomes. Note that  $\tau(x^{l}) =   \mathbb{E}[\beta(X^{l}, e) | X^{l} = x^{l}]$ can always be estimated using a local linear regression of $\beta(X^{l}, e)$ on $X^{l}$, regardless of the type of outcome. 
However, when the PSR is used for a discrete outcome, particular care is needed for the potential model extrapolation when estimating $\beta(X^{l}, e)$. 
 This is because in the context of discrete outcomes, the estimation of the model (\ref{beta working}), 
$
Y  =   \beta(X^{l}, e)\cdot D  +   \E{ Y(0) | X^{l}, e }  + \xi$,    
is often done separately for each treatment group, instead of two treatment groups together. Note that $\beta(X^{l}, e)$ can be rewritten with two components,
\[  
\beta(X^{l}, e) = \E{Y(1) - Y(0) | X^{l}, e} = \E{Y(1) | X^{l}, e}  - \E{Y(0) | X^{l}, e}. 
\]
As such, alternative methods are needed to estimate $\beta(X^{l}, e)$ without separating the two treatment groups.  This is left to future work. 
\section*{Supplementary Material} 
The online Supplementary Material includes technical proofs, 
 additional numerical results from the simulation study and empirical application, and extensions of the proposed method. 
%  to cases of multidimensional covariate $x^{l}$ and discrete $x^{l}$. 

%%%%%%%%%%%%%%%%%%%%%%%%%%%%%%%%%%%%%%%%%%%%%%%%%%%%%%%%%%%%%%%%%%%%%%%%%%%%%%%%%%%%%%%%%%%%%%%%%%%%%%%%%%%%%%%%%%%%%%%%%%%%
\section*{Acknowledgments}
The authors thank the assistant editor and anonymous reviewers for their helpful comments and valuable suggestions. 
This research was supported by the National Natural Science Foundation of China (No. 11971064, 12071015, and 12171374) and the Major Project of National Statistical Science Foundation of China (No. 2021LD01). 

%%%%%%%%%%%%%%%%%%%%%%%%%%%%%%%%%%%%%%%%%%%%%%%%%%%%%%%%%%%%%%%%%%%%%%%%%%%%%%%%%%%%%%%%%%

\begin{spacing}{1.25}

\bibhang=1.7pc
\bibsep=2pt
\fontsize{9}{14pt plus.8pt minus .6pt}\selectfont
\renewcommand\bibname{\large \bf References}
%\begin{thebibliography}{11}
\expandafter\ifx\csname
natexlab\endcsname\relax\def\natexlab#1{#1}\fi
\expandafter\ifx\csname url\endcsname\relax
  \def\url#1{\texttt{#1}}\fi
\expandafter\ifx\csname urlprefix\endcsname\relax\def\urlprefix{URL}\fi

%% use bibfile 
  \bibliographystyle{chicago}      % Chicago style, author-year citations
  \bibliography{bib}   % name your BibTeX data base

\end{spacing}

%%  Another method

%%%%%%%%%%%%%%%%%%%%%%%%%%%%%%%%%%%%%%%%%%%%%%%%%%%%%%%%%%%%%%%%%%%%%%%%%%%%%%%%%%%%%%%%%%%%%%%%%%%%%%%%%%%%%%%%%%%%%%%%%%%%
%\vskip .65cm
%\noindent
%first author affiliation
%\vskip 2pt
%\noindent
%E-mail: (first author email)
%\vskip 2pt
%
%\noindent
%second author affiliation
%\vskip 2pt
%\noindent
%E-mail: (second author email)

% \vskip .3cm
%\centerline{(Received ???? 20??; accepted ???? 20??)}\par
\end{document}